# Comments on 'Frequency Diverse Array Antenna Using Time-Modulated Optimized Frequency Offset to Obtain Time-Invariant Spatial Fine Focusing Beampattern'

M. Fartookzadeh

In the recent papers [1-4] including above paper time modulated frequency diverse arrays (FDAs) have been presented to obtain time invariant spatial patterns. The presented FDAs in [1-3] have the feature of time-invariant spatial focusing which means they have a constant maximum in a time duration, $T$, at a point with desired range, $r$, and angle, $\theta$. In [4] the pattern is time invariant, yet, not focused. However, in this communication it is indicated that the patterns are obtained using incorrect definition of time in some equations. The equation system of [3] is explained here that can be extended to [1, 2, 4], explicitly.

For the antenna array with $(2N+1)$ elements, each labeled by an integer number, $n$ ($n \in [-N,N]$), the array factor (AF) is given by [3, 5]

$$AF(t,r,\theta) = \sum_{n=-N}^{N} e^{j\left\{2\pi\left[\Delta f_n\left(t-\frac{r}{c}\right)+\frac{nd\sin\theta}{c}(f_0+\Delta f_n)\right]+\phi_n\right\}}, \quad (1)$$

where $f_0$ is frequency of the center element and $(f_0 + \Delta f_n)$ is the frequency of $n$-th element with the assumptions, $\Delta f_n = \Delta f_{-n}$ and $\Delta f_0 = 0$. $c$ is the speed of electromagnetic (EM) wave propagation and $d$ is the spacing between elements. $\phi_n$ is a constant phase difference corresponding $n$-th elements. The AF as a function of time, $t$, indicates a wave moving with velocity, $c$, while for $\Delta t$ time shifting $c\Delta t$ movement in range, $r$, is required to obtain the same AF. Now, the practicability of compensating the wave speed by defining each $\Delta f_n$ as a function of time is discussed in the following explanations.

It is obvious that each $\Delta f_n$ can be defined as a function of time at the position of $n$-th element. However, relocating from its position, at the same time, the observed $\Delta f_n$ is for the past. For example, in the distance $\Delta r$ from the array, the observed $\Delta f_n$ is for $\Delta r/c$ before the present time. Therefore, $\Delta f_n$ in the AF cannot be a function of time and the propagation delay should be considered. In particular, each $\Delta f_n$ can be defined as a function of $(t_0 = t - r/c)$. This is while in [1-4] the observed frequency differences are defined as the functions of time in the AF. This assumption means that the frequency of each element is a function of time and range, which is not feasible. Consequently, in Eq. (3) of [3], $\Delta f_n(t)$ should be replaced by $\Delta f_n(t - r/c)$ (admitting that $r \gg d$) and therefore it is $\Delta f_n(t - r/c)$ that should be set to Eq. (6), leading to a dependence of $\Delta f_n$ on $r$. The same applies to Eq. (12) and (13) of [1].

An easy-understanding evidence for observing incorrectness of this assumption is the power pattern plots in [1-4]. For example, it can be observed in Fig. 8(a) of [3] that the maximum of AF is obtained in the range $r_1 = 15$ m, at $t = 0$ using the frequency offsets

$$\Delta f_n = \frac{g_n - \frac{n}{2}\sin\theta_0}{t - \frac{r_1}{c} + \frac{n}{2f_0}\sin\theta_0}, \quad t \in [0,T], \quad (2)$$

where $\theta_0 = -30°$ is the angle of focus, $g_n$ is an optimized factor for $n$-th element and $T = 30$ ns. This EM power should have been sent from the antenna array at 50 ns before; if we agree that the EM wave travels in space-time on light-cone with the velocity $c = 3 \times 10^8$ m/s. It can be observed that there is no information from the frequencies of the elements in the past and the excitation has been started at $t = 0$. Therefore, the power could not have appeared at the same time to the range, $r_1 = 15$ m.

Nevertheless, it is worth noting that the optimization method in [1-3] is interesting and useful. For example, similar spatial power pattern with Fig. 6 of [3] can be obtained for the FDA with $N = 5$, $f_0 = 3$ GHz, and $d = \frac{1}{2}c/f_0$ at $t = 0$ using the same optimized factors, $([g_n] = [g_{-n}] = [1.8, 4.4, 4.4, 5.5, 4.8]$, reading from Fig. 4 of [3]) as indicated in Fig. 1(a) by removing $t$ from $\Delta f_n$ in (2). However, at $t = T$ the power pattern will be changed and the focus point will be at $r = 24$ m as indicated in Fig. 1(b). Projection of power-pattern on the time-range axes for the constant $\Delta f_n$ is indicated in Fig. 1(c).

It should be noted that the maximum power is observed again at $t = 0$ on $r = 15$ m for the constant $\Delta f_n$ as can be seen in Fig. 1(c). The zero time is conventional, while the frequencies of propagated signals of the elements are constant with time. In fact, the propagation has begun from the earlier time for constructing this power pattern. Therefore, the resulting power pattern at this range is due to excitation in the past. For example, if the excitation began from $t = -90$ ns the power pattern would be as indicated in Fig. 2(a). However, if it is desired to begin the excitation at $t = 0$, one can change the frequency offsets as

$$\Delta f_n = \frac{g_n - \frac{n}{2}\sin\theta_0}{t_m + \frac{n}{2f_0}\sin\theta_0}, \quad (3)$$

where $t_m$ is the desired time of appearing the maximum power at the location of array. In the previous case $t_m$ was in fact -50 ns. The power pattern of a similar FDA with $t_m = 40$ ns is indicated in Fig 2(b), assuming that the excitation has begun from $t = 0$.

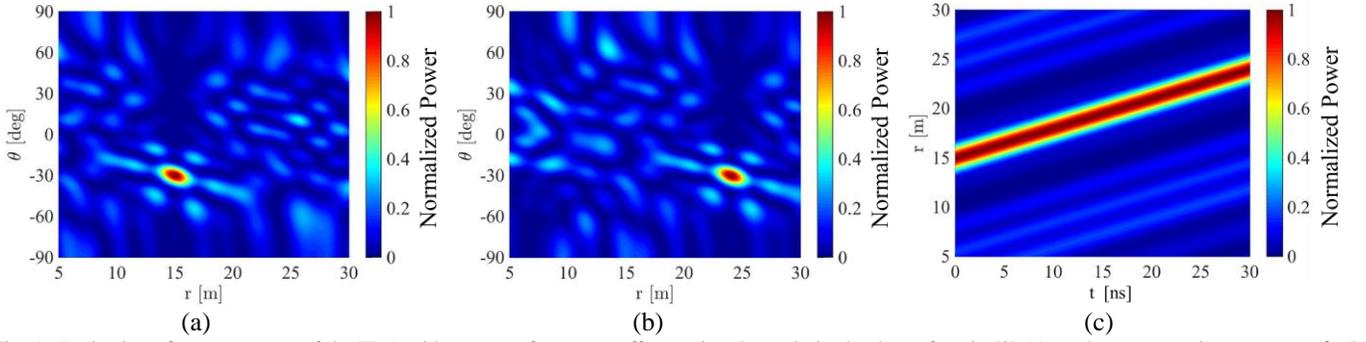

Fig. 1. Projection of power pattern of the FDA with constant frequency offsets using the optimized values of $g_n$ in [3] (a) on the range-angle axes at $t = 0$, (b) on the range-angle axes at $t = 30$ ns and (c) on the time-range axes at $\theta = -30°$.

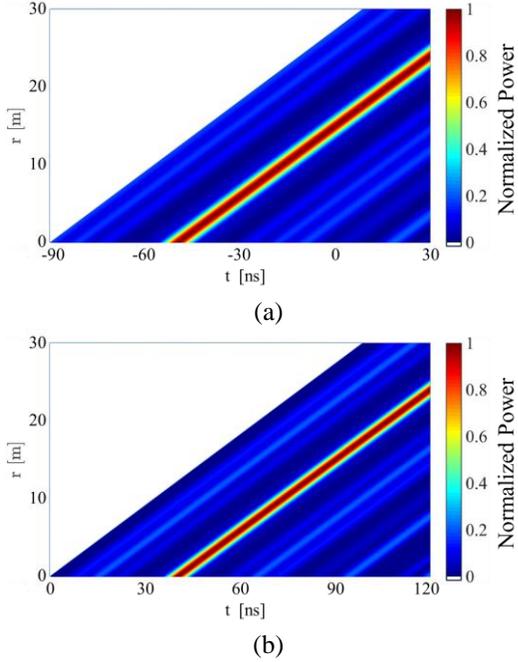

Fig. 2. Projection of power pattern of FDA with constant frequency offsets using the optimized values of $g_n$ in [3] on the time-range axes at $\theta = -30°$; (a) similar power pattern with Fig. 1(c) when the excitation begins from $t = -90$ ns, (b) the power pattern when $\Delta f_n$ is defined by (3), $t_m = 40$ ns, and the beginning of excitation is at $t = 0$.

In addition, constant range for the focus point of FDAs with $r_1 \gg d$, appears to be unattainable using any method [6-8], since the velocity of EM wave is independent from its frequency, amplitude, etc. Thus, the power delivered by the array to the range, $r_1$, will appear in the range $r_1 + c\Delta t$ after elapsing the time, $\Delta t$. Consequently, the focusing point of antennas always moves with the velocity of wave propagation in farfield. However, constant nearfield focusing antennas are accessible [9-11], since the nearfield range is not similar for elements of an array or parts of an antenna. Furthermore, it is not in contradiction with movement of the EM wave in space-time on the light-cone, since the EM power has been reached to the focusing point from different locations.